\documentclass[pre, twocolumn, showpacs,superscriptaddress]{revtex4}
\usepackage{amsmath,amssymb}
\usepackage{graphicx}
\usepackage{pslatex}

\begin{document}

\def\be{\begin{equation}}
\def\ee{\end{equation}}
\def\bfi{\begin{figure}}
\def\efi{\end{figure}}
\def\bea{\begin{eqnarray}}
\def\eea{\end{eqnarray}}

\title{Self tuning phase separation in a model with competing interactions inspired by biological cell polarization }

\author{T. Ferraro}
\email{ferraro@na.infn.it}
\affiliation{Dipartimento di Scienze Fisiche, Universit\`a di
  Napoli ``Federico II'',\\ Complesso Universitario di Monte
  Sant'Angelo, via Cintia 80126 Napoli, Italy}
\affiliation{INFM-CNR Coherentia}
\author{A. Coniglio}
\affiliation{Dipartimento di Scienze Fisiche, Universit\`a di
  Napoli ``Federico II'',\\ Complesso Universitario di Monte
  Sant'Angelo, via Cintia 80126 Napoli, Italy}
\affiliation{INFM-CNR Coherentia}
\affiliation{INFN Udr di Napoli}
\author{M. Zannetti}
\affiliation{Dipartimento di Matematica ed Informatica, Universit\`a di Salerno,\\ via Ponte don Melillo, 84084 Fisciano (SA), Italy}


\begin{abstract}
We present a theoretical study of a system with competing
short-range ferromagnetic attraction and a long-range
anti-ferromagnetic repulsion, in the presence of a uniform
external magnetic field. The interplay between these interactions,
at sufficiently low temperature, leads to the self-tuning of the
magnetization to a value which triggers phase coexistence, even in
the presence of the external field. The investigation of this
phenomenon is performed using a Ginzburg-Landau functional in the
limit of an infinite number of order parameter components (large
$N$ model). The scalar version of the model is expected to
describe the phase separation taking place on a cell surface when
this is immersed in a uniform concentration of chemical stimulant.
A phase diagram is obtained as function of the external field and
the intensity of the long-range repulsion. The time evolution of
order parameter and of the structure factor in a relaxation
process are studied in different regions of the phase diagram.
\end{abstract}

\maketitle

\section{Introduction}
Several natural and social processes are governed by competing interactions.
Often the interplay between opposite actions produces ordered phases and symmetry breaking
events.
For example, models based on competing interactions are able to explain lamellar
phases in charged colloids \cite{TCpre2007}, pattern formation in magnetic films,
Languimir monolayers and liquid crystals \cite{Giuliani} and some market behaviors \cite{econofis}.
In living organisms there is an important example of these processes: the spatial orientation
of eukaryotic cells \cite{alberts} called eukaryotic directional sensing.
Many eukaryotic cells are able to orient (polarize) for moving along directions after
an external stimulation. This process is fundamental for important biological functions
like morphogenesis of organs and tissues, wound healing, immune response, social behaviors of some amoeboid cells.
The process of orientation takes place on the cell membrane where
the pattern formation of
domains of two different enzymes determines a symmetry breaking
which triggers the directional sensing \cite{JMD+04}.
Pattern formation occurs as a response to an external
stimulation, usually a chemical signal activating
specific receptors on the cell surface, enhanced by
a cascade of chemical reactions leading
to the cell polarization (see Ref. \cite{GCT+05} and references therein).
Experimental
observations \cite{PRG+04} suggest that the domain formation is a consequence
of self-organization of molecular patches.

Let us briefly summarize the biological mechanism of directional
sensing. It can be explained in term of the interplay between two
enzymes: PTEN (phosphatase and tesin homolog) and PI3K
(phosphatidylinotisol 3-kinase). This interplay is mediated by two
lipids: the PIP$_2$ (phosphatidylinotisol bisphosphate) and the
PIP$_3$ (phosphatidylinotisol trisphosphate). Before stimulation,
the cell membrane is populated only by the PTEN enzyme with its
product PIP$_2$ but, when the external chemoattractant is switched
on, the enzyme PI3K goes from cytoplasm to the cell membrane and
binds to receptors. Then, the interplay between the two enzymes
takes place: the PI3K catalyzes PIP$_2$ in PIP$_3$ and PTEN
catalyzes PIP$_3$ in PIP$_2$. The enzymes can bind to the
respective lipid products which diffuse over the membrane. Enzymes
can unbind from membrane and quickly diffuse in cytoplasm binding
again in another place of the membrane. Catalysis and lipides
diffusion mediate an effective short-range attraction between
enzymes of the same type. The quick diffusion of enzymes in the
cytoplasm mediates a long range interaction. The combination of
these actions produces the phase separation of a PI3K rich zone
and a PTEN rich zone.

The natural framework to treat the process, from a physical point
of view, is the statistical physics of phase separation, which is
useful to understand and describe in synthetic way the behavior of
many complex systems that give rise to organization and pattern
formation \cite{Domb, preSagDes1994}. The spatial organization
phenomena described above have the characteristic of
self-organized phase separation processes, where the cell state,
driven by an external field, decays into the coexistence of two
chemical phases, spatially localized in different regions. It was
shown recently, by Monte Carlo simulation in a lattice gas model
\cite{epl+FDGC+08}, that the phenomenology of directional sensing
can be obtained by using an effective free energy. A similar point
of view was used in the recent papers of Gamba et al.
\cite{GKL+07,GKLO+08}. The remarkable physical characteristic of
the process is that the orientation is possible for a wide range
of external chemical attractant \cite{Zig77}. Namely, the phase
coexistence and separation is possible for different values of an
external field. In the lattice gas model \cite{epl+FDGC+08}, a
long range repulsion, derived from the interaction with a finite
cytosolic reservoir of total enzymes and the interaction with an
external field, modelling the action of the chemical attractant,
give rise to coexistence for an interval of values of the external
field. A short-range attraction between enzymes, derived from
their catalytic actions on lipids and diffusion, gives rise to a
coarsening process which produces phase separation. A similar
mechanism operates in some econophysics models \cite{econofis}
based on two major conflicting interactions in economy, the
tendency of a trader to follow the actions of his neighbors and
the tendency to follow the actions of the minority. In the
language of magnetic systems, which we will use throughout this
paper, the first tendency can be modelled by a short-range
ferromagnetic interaction, while the second one by an
antiferromagnetic long range interaction.

Inspired by the mechanism of eukaryotic directional sensing and motivated by its wide applicability,
here we present the analytical treatment of a system with competing
short-range attraction and long-range repulsion, under the action of a uniform external field.
This is done in the framework of the time dependent Ginzburg-Landau (TDGL) theory for the evolution
of the order parameter. The analytical tractability of the dynamics is achieved considering
a vector order parameter in the limit of an infinite number of components (large $N$ limit) \cite{preCRZ1994}.
The equations of motion are derived in the scheme of non conserved order
parameter, corresponding to the absence of local concentration of enzymes.

The large $N$ limit is a powerful method to obtain analytical
results, nevertheless it is necessary to keep in mind some
important differences with the nonlinear models usually employed
in the description of phase ordering when the order parameter is a
scalar, such as kinetic Ising models or the TDGL with $\phi^4$
interaction~\cite{Bra95}. The basic difference is in the mechanism
of equilibration after a symmetric quench below the critical
point. In the nonlinear models the system responds to the
dynamical instability by the formation and growth, through
coarsening, of domains of the ordered phases. This leads to the
development of a bimodal probability distribution for the local
magnetization, which eventually ought to evolve into the symmetric
mixture of the two possible broken symmetry ordered
phases~\cite{footnote}. This we call an {\it ordering process}. In
the large $N$ limit, instead, the development of a bimodal
distribution, and therefore ordering, is not possible, since, as
it will be clear below, the system is effectively linearized and
the statistics are Gaussian. Then, the response to the dynamical
instability takes place through the development of macroscopic
fluctuations in the most unstable Fourier component of the order
parameter, through a process which is formally identical to the
one leading to the formation of the condensate in the low
temperature phase of the ideal Bose gas. We refer to this
equilibration mechanism as {\it condensation of fluctuations}. The
differences between the two equilibration processes have been
investigated in detail in Ref.~\cite{CCZ,Fusco}.

The remarkable feature of the large $N$ limit, and the reason for
its wide use, is that despite the considerable difference in the
physical processes of equilibration, the phenomenology of the
observables  of interest, such as correlation functions and
response functions, is the same as in the nonlinear models, apart
for obvious quantitative discrepancies, like the values of
exponents or the shape of scaling functions. The typical example
is that of the equal time structure factor, which displays
dynamical scaling and the growth of the Bragg peak in the large
$N$ limit~\cite{CZ}, exactly as in the nonlinear models. It is,
then, matter of interpretation in one case to read the growth of
the Bragg peak as revealing domain coarsening and, in the other,
condensation of fluctuations. There is, by now, a vast body of
literature documenting the robustness of the large $N$ limit in
reproducing the phenomenology of phase ordering in a large variety
of models, warranting to overlook the distinction between ordering
and condensation, as we shall do in this paper, when one is
interested in the main qualitative features of the process.

The paper is organized as follows. In section II we carry out the
large $N$ limit on the TDGL model for a vector order parameter,
deriving the basic equations. In section III we study the
equilibrium properties of the system, obtaining the phase diagram
in the temperature and external field plane, parameterized by the
strength of the long-range repulsion. This delimits the region of
parameters where condensation or, equivalently, phase coexistence
and separation are possible. Section IV is devoted to the study
the time behavior of the average value of the order parameter
(magnetization) and of the correlation function. We recall that in
the scalar case the magnetization represents the concentration
difference between the two species of enzymes. Concluding remarks
are presented in section V.

\section{The model}

The system is modelled by a free energy functional of the form
\begin{eqnarray}
{\cal H}[\vec{\phi}] & = & \int_V d \vec{x} \left \{ {1 \over 2}(\nabla\vec{\phi})^2 +
{r \over 2}(\vec{\phi}\cdot \vec{\phi})  + {g \over 4 N}(\vec{\phi}\cdot \vec{\phi})^2 -
\vec{H}(\vec{x})\cdot \vec{\phi}(\vec{x})  \right \}
\nonumber \\
& + & {1 \over 2}{\lambda \over V}\left [ \int_V d \vec{x} \, \vec{\phi} \right ]^2
\label{1.1}
\end{eqnarray}
where $\vec{\phi} = (\phi_1,...,\phi_N)$ is an $N$ component vector order parameter.
We shall take  $r<0, \, g>0, \, H \sim {\cal O}(N^{1/2})$,
$V$ is the volume of the system and $\lambda>0$ is an antiferromagnetic coupling.
We shall consider the static and dynamic properties of the model.

As is well known from the theory of critical phenomena, the
introduction of an $N$ component order parameter is a very
convenient technical device to generate controlled and systematic
correction about mean field behavior, using $1/N$ as an expansion
parameter~\cite{ZJ}. We shall limit the treatment to the lowest
(mean field) order by taking the large $N$ limit ($N \rightarrow
\infty$).

\subsection{Equation of motion}

In the framework of TDGL model for
the dynamics
\be
{\partial \vec{\phi}(\vec x,t) \over \partial t}  =
- {\delta {\cal H}[\vec{\phi}] \over \delta \vec{\phi}}(\vec x,t)
+ \vec{\eta}(\vec x,t)
\label{EM1}
\ee
where $\vec{\eta}(\vec x,t)$ is the white noise at temperature $T$, with zero average and correlator
\[
\langle \eta_{\alpha}(\vec x,t) \eta_{\beta}(\vec x^{\prime},t^{\prime})\rangle =
2T\delta_{\alpha\beta}\delta(\vec x - \vec x^{\prime})\delta(t - t^{\prime})
\]
the equation of motion of the order parameter is given by
\begin{eqnarray}
{\partial \vec{\phi}(\vec x,t) \over \partial t} & = &
 - \left [-\nabla^2\vec{\phi}(\vec x,t) +r\,\vec{\phi}(\vec x,t) +
{g \over N}(\vec{\phi}\cdot \vec{\phi})\vec{\phi}(\vec x,t) -\vec{H}(\vec x) \right ]
\nonumber \\
& - & {\lambda \over V} \int_V d \vec{x} \, \vec{\phi}(\vec x,t) + \vec{\eta}(\vec x,t).
\label{1.2}
\end{eqnarray}
Since the external field breaks the rotational symmetry in the order parameter space, it is convenient
to introduce the longitudinal and  transverse components with respect $\vec{H}(\vec{x})$
\be
\vec{\phi}  =  \vec{\phi}_{\parallel} + \vec{\phi}_{\perp}
\label{1.3}
\ee
and then to split the longitudinal component into the sum
\be
\vec{\phi}_{\parallel}(\vec x,t) = \vec{M}(\vec x,t) + \vec{\psi}(\vec x,t)
\label{1.5}
\ee
where $\vec{M}(\vec x,t) = \langle \vec{\phi}_{\parallel}(\vec x,t) \rangle$
is the magnetization, while the average longitudinal fluctuations vanish
$\langle \vec{\psi}(\vec x,t) \rangle \equiv 0$ by construction. The angular brackets
denote the average over both the initial condition and the thermal noise.
In the following we shall take a reference frame with the $1$-axis along the longitudinal direction.

Assuming, next, $M \sim {\cal O}(N^{1/2})$, $\psi \sim {\cal O}(1)$ and comparing terms
of the same order of magnitude in $N$,
to leading order we get the pair of equations
\be
{\partial m \over \partial t} = - \left [ -\nabla^2 + r  + g m^2 +
g S \right ] m + \big (h - {\lambda \over V}\int_V d \vec{x} \, m \big )
\label{1.8bis}
\ee
and
\be
{\partial  \vec{\phi}_{\perp} \over \partial t} = - \left [ \Big (-\nabla^2 + r +
g  m^2 + gS  \Big ) \vec{\phi}_{\perp}
+ {\lambda \over V }\int_V d \vec{x} \, \vec{\phi}_{\perp}\right ]
+ \vec{\eta}_{\perp}
\label{1.10bis}
\ee
where $m \equiv m(x,t)$ and $h \equiv h(x,t)$ are the following rescaled quantities
\be
m(\vec x,t) = M(\vec x,t)/N^{1/2},\;\;h(\vec x,t) = H(\vec x,t)/N^{1/2}.
\label{redvar}
\ee
In the large $N$ limit the quantity $S(\vec x,t)$ is given by the self-averaged fluctuations
\be
\lim_{N \to \infty}{1 \over N}(\vec{\phi}_{\perp}\cdot\vec{\phi}_{\perp}) =
\langle \phi_{\alpha} \phi_{\alpha} \rangle= S(\vec x,t)
\label{1.11}
\ee
of the generic transverse component $\phi_{\alpha}$. Furthermore,
since in Eq.~(\ref{1.10bis}) the components of $\vec{\phi}_{\perp}$ are effectively decoupled,
from now on we shall refer to the equation for the generic component omitting
the $\alpha$ subscript.

Taking a space and time independent external field $h$ and space
translation invariant initial conditions, we can assume space
translation invariance to hold at all times. Hence, Fourier
transforming with respect to space, and introducing the equal-time
transverse structure factor \be \langle
\phi(\vec{k},t)\phi(\vec{k}^{\prime},t) \rangle =
C_{\perp}(\vec{k},t) V\delta_{\vec{k}+\vec{k}^{\prime},0}
\label{trcf} \ee we obtain the closed set of equations \be
{\partial m(t)  \over \partial t} = - \omega(0,t)m(t) + h
\label{1.12bis} \ee \be {\partial C_{\perp}(\vec{k},t)\over
\partial t} = -2 \omega(k,t)C_{\perp}(\vec{k},t) + 2T \label{1.18}
\ee \be S(t) = {1 \over V}\sum_{\vec{k}} C_{\perp}(\vec{k},t)
\label{1.19} \ee with $m(t) \equiv m(\vec k=0,t)$ and
$\omega(k,t)$ defined by \be \omega(k,t)= k^2 + \lambda
\delta_{k,0} + r + g (m^2 + S) \label{1.14} \ee and the noise
correlator in Fourier space given by \be \langle \eta(\vec{k},t)
\eta(\vec{k}^{\prime},t^{\prime})\rangle =
2TV\delta_{\vec{k}+\vec{k}^{\prime},0}\delta(t-t^{\prime}).
\nonumber \label{1.16} \ee With periodic boundary conditions the
wave vector runs over $\vec k = {2\pi \over L}\vec n$, where $\vec
n$ is a vector with integer components and $L^d=V$. Furthermore,
sums over $\vec k$ like the one in Eq.~(\ref{1.19}) are cutoff to
the upper value $k_{max}=\Lambda$, where $\Lambda^{-1}$ is related
to a characteristic microscopic length, for instance the lattice
spacing of an underlying lattice. Finally, the longitudinal
fluctuations $\psi$ have been dropped since do not give any
contribution to leading order.

\section{Static properties and phase diagram}

If equilibrium is reached, all quantities become time independent. Rewriting Eq.~(\ref{1.14}) as
\be
    \omega(k) = \left \{ \begin{array}{ll}
        \lambda + \mu , \qquad $for$ \qquad k=0 \\
        k^2 + \mu,  \qquad $for$ \qquad  k \neq 0
        \end{array}
        \right .
        \label{Eq.1}
        \ee
with
\be
\mu = r + g (m^2 + S)
\label{Eq.2}
\ee
and putting to zero the time derivatives, from Eqs.~(\ref{1.12bis}), (\ref{1.18}) and~(\ref{1.19})
we obtain the set of equations
\be
(\lambda + \mu) m = h
\label{1.12tris}
\ee
\be
\omega(k)C_{\perp}(\vec{k}) = T
\label{1.18tris}
\ee
\be
S = {1 \over V}\sum_{\vec{k}} C_{\perp}(\vec{k}).
\label{1.19tris}
\ee
From Eqs.~(\ref{trcf}) and~(\ref{1.18tris})   follows $C_{\perp}(\vec{k}) \geq 0$ and
$\omega(k) \geq 0$, respectively. The latter inequality, because of Eq.~(\ref{Eq.1}), requires
\be
\mu \geq \mu_{min} = -k^2_{min}
\label{mu.1}
\ee
where $k_{min} \sim 1/L$ is the minimum allowed value of
$k \neq 0$.
Therefore, for a given $\lambda$ and for $V$ sufficiently large, $\lambda + \mu >0$ and
Eq.~(\ref{1.12tris}) can be rewritten as
\be
m={h \over \lambda + \mu}.
\label{Eq.m}
\ee
In the same way, from Eq.~(\ref{1.18tris}) we can write
\be
    C_{\perp}(\vec{k}) = \left \{ \begin{array}{ll}
        T /(\lambda + \mu) , \qquad $for$ \qquad k=0 \\
        T /(k^2 + \mu),  \qquad $for$ \qquad  k \geq k_{min}
        \end{array}
        \right .
        \label{Eq.3}
        \ee
where $C_{\perp}(\vec{k}_{min})$ diverges as  $\mu$ approaches $\mu_{min}$.
Notice that $\mu$ can be identified with the inverse square
transverse correlation length $\xi^{-2}$.

In order to obtain the full solution, we must now determine how
$\mu$ depends on the parameters of the problem $(T,h,V,\lambda)$.
In principle, this can be done by inserting the above results into
Eq.~(\ref{Eq.2}) and solving the basic self-consistency equation
\be \mu -  g \left ({h \over \lambda + \mu} \right )^2   = r +  {g
\over V} {T \over \lambda + \mu} +  {Tg \over V}\sum_{\vec{k} \neq
0}{1 \over k^2 + \mu}. \label{Eq.4} \ee However, for our purposes
general considerations are sufficient, without actually solving
the above equation. For $V$ sufficiently large the second term in
the right hand side can be neglected and the equation can be
rewritten in the form \be \mu -  g \left ({h \over \lambda + \mu}
\right )^2   = r +  {g \over V} {T \over \mu - \mu_{min}} +  {Tg
\over V}\sum_{\vec{k} > \vec{k}_{min}}{1 \over k^2 + \mu}
\label{Eq.4bis} \ee where the $k_{min}$ term has been extracted
from underneath the sum. Letting $\mu$ to vary from $\mu_{min}$ to
$\infty$, the left hand side is a monotonously increasing function
of $\mu$, while the right hand side diverges at $\mu_{min}$ and
decreases  monotonously with increasing $\mu$. Therefore, for any
finite $V$, there exists a solution $\mu^*(V) > \mu_{min}$.
Looking at Eqs.~(\ref{Eq.m}) and~(\ref{Eq.3}), this means that the
system behaves paramagnetic  all over the $(T,h)$ plane, with a
finite structure factor. The difference with respect to what one
would have in the purely short-range model, due to $\lambda \neq
0$, is revealed by the anomaly~(\ref{Eq.3}) in the structure
factor at $k=0$ and by the reduction of the magnetization in
Eq.~(\ref{Eq.m}). Rewriting the latter as $\mu^*(V)m = h_{eff}$
with \be h_{eff} =h -\lambda m, \label{Eq.5} \ee we see that the
reduction of the magnetization comes about through a feedback
mechanism, whereby the external field $h$, via the long-range
interaction, is substituted by $h_{eff}$.

\begin{figure}
\begin{center}
  \includegraphics[width=75mm]{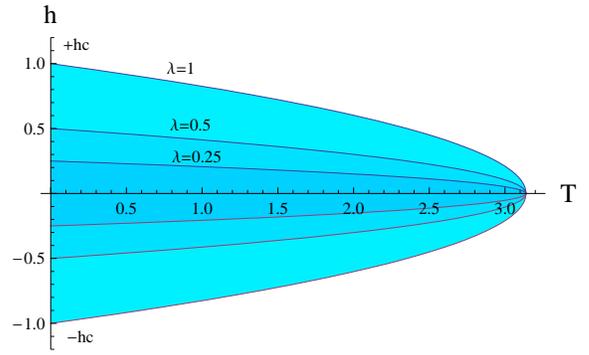}
  \caption{(Color online) Phase diagram showing the $\lambda$ dependence of the region within which
    condensation of the transverse fluctuation takes place.
    The presence of a condensate, in the blue regions, corresponds to phase separation and
    coexistence in the corresponding scalar model.
    In this and in all the other figures
    $g=-r=1$, $\Lambda=2\pi$ and $d=3$, yielding $T_C(0)= \pi$.
  }
  \label{phase}
\end{center}
\end{figure}

Let us now see what happens in the infinite volume limit. From Eq.~(\ref{mu.1})
follows $\mu_{min} \rightarrow 0^-$ and there are two possibilities
\be
    \lim_{V \to \infty} \mu^*(V)= \left \{ \begin{array}{ll}
        \mu^*  > 0 \\
        \mu^*  = 0.
        \end{array}
        \right .
        \label{Eq.10}
        \ee
In the first case, the second term in the right hand side of Eq.~(\ref{Eq.4bis}) can be neglected yielding
\be
\mu -  g \left ({h \over \lambda + \mu} \right )^2   = r +  TgB(\mu)
\label{Eq.11}
\ee
where
\be
B(\mu) = \lim_{V \to \infty} {1 \over V}\sum_{\vec{k} \neq 0}{1 \over k^2 + \mu}
= K_d \int_0^{\Lambda} dk \, {k^{d-1} \over k^2 + \mu}
\label{Eq.7}
\ee
is a monotonously decreasing function of $\mu$ with the maximum value
\be
    B(0)= \left \{ \begin{array}{ll}
        K_d {\Lambda^{d-2} \over d-2},\qquad $for$ \qquad d>2 \\
        $divergent for$ \qquad d \leq 2
        \end{array}
        \right .
        \label{Eq.8}
        \ee
with $K_d=[2^{d-1}\pi^{d/2}\Gamma(d/2)]^{-1}$ being the $d$-dimensional solid angle~\cite{nota1}
and $\Gamma(d/2)$ the Euler gamma function.
Hence, Eq.~(\ref{Eq.11}) admits a positive solution if $T$ is greater than
the $h$-dependent critical temperature
\be
T_C(h) = -{r + gh^2 /\lambda^2 \over g B(0)}
\label{2.10}
\ee
which vanishes for $d \leq 2$ and for any $h$, while it is finite
for $d >2$ reaching the maximum value $T_C(0)= -r/g B(0)$
for $h=0$ and decreasing to zero when $h$ reaches the limit values $\pm h_C$ (see Fig. \ref{phase})
with
\be
h_C = \lambda(-r/g)^{1/2}.
\label{2.11}
\ee

Conversely, if $T$ and $h$ are such that $T < T_C(h)$, Eq.~(\ref{Eq.11}) cannot be satisfied.
This means that the second of the two possibilities in Eq.~(\ref{Eq.10}) applies,
requiring to keep also the second term in the right hand side of Eq.~(\ref{Eq.4bis}),
which we rewrite as
\be
\mu -  g \left ({h \over \lambda + \mu} \right )^2   = r +  {g \over V} C_{\perp}(\vec{k}_{min})
+  Tg B(\mu).
\label{Eq.12}
\ee
Since this is satisfied for $\mu=0$, we get
\be
C_{\perp}(\vec{k}_{min}) = V \left [ -{r \over g} - \left ({h \over \lambda} \right )^2 \right ]
\left [ {T_C(h) - T \over T_C(h)} \right ]
\label{Eq.13}
\ee
showing that $C_{\perp}(\vec{k}_{min})$, for $T < T_C(h)$, diverges like the volume in order
to give a finite contribution in the right hand side of Eq.~(\ref{Eq.12}).

Summarizing, in the infinite volume limit \be
    m = \left \{ \begin{array}{ll}
        h /(\lambda + \mu^*), \qquad $for$ \qquad T > T_C(h) \\
        h /\lambda,  \qquad $for$ \qquad  T < T_C(h)
        \end{array}
        \right .
        \label{2.15}
        \ee
where $\mu^*>0$, and
\be
    C_{\perp}(\vec{k})  = \left \{ \begin{array}{ll}
        T / (k^2 + \mu^*), \qquad $for$ \qquad T > T_C(h) \\
        T / k^2  + {\cal M}^2(T)\delta(\vec{k}-0^+),  \qquad $for$ \qquad  T < T_C(h)
        \end{array}
        \right .
        \label{Eq.16}
        \ee
showing that there is condensation of transverse fluctuation at $k=0^+$
for $T < T_C(h)$, with the size of the condensate given by
\be
{\cal M}^2(T) = \left [ -{r \over g} - \left ({h \over \lambda} \right )^2 \right ]
\left [ {T_C(h) - T \over T_C(h)} \right ].
\label{Eq.18}
\ee

In order to understand this result, it should be recalled that in
the purely short-range large $N$ model the phase transition occurs
only on the $h=0$ axis, where for $T < T_C(0)$ the condensation of
fluctuations takes place~\cite{CCZ} at $k=0$. Condensation of
fluctuations means that $ C_{\perp}(\vec{k}=0)$ becomes
macroscopic in order to equilibrate the system below $T_C(0)$
without breaking the symmetry, through a mechanism very similar to
that of the Bose-Einstein condensation, as mentioned in the
Introduction. No other mechanism is available, since the large $N$
limit renders the system effectively Gaussian~\cite{CCZ,Fusco}.
However, condensation of fluctuations produces the onset of a
Bragg peak at $\vec{k}=0$ and, therefore, a phenomenology of the
structure factor which is indistinguishable from that due to the
occurrence of phase separation in the nonlinear
models~\cite{Bra95}. When $h \neq 0$ the symmetry is broken and
equilibrium can be established at any temperature through the
development of a non vanishing magnetization, without any
condensation.

In the system with the long-range coupling of antiferromagnetic
type everything remains the same along the $h=0$ axis, since the
symmetry is unbroken, the magnetization is zero and the only
effect of the $\lambda$ term is to shift the Bragg peak from $\vec
k=0$ to $\vec k=0^+$. The novelty appears outside of the $h=0$
axis, where the explicitly broken symmetry induces the development
of a non vanishing magnetization which, however, through the
feedback mechanism driven by the antiferromagnetic interaction
produces the effective reduction~(\ref{Eq.5}) of the external
field. So, if for a given $\lambda$, the values of $T$ and $h$
manage to make $h_{eff}=0$, at that point the value of the
magnetization gets stabilized to the value~(\ref{2.15}) and the
only way to equilibrate the system is through the condensation of
fluctuations. The result is the phase diagram of Fig. \ref{phase},
showing the expansion of the phase coexistence region outside the
$h=0$ axis. The constant $\lambda$ curves delimit the regions on
the $(T,h)$ plane within which the system self-tunes the final
magnetization to the value $h/\lambda$ such that $h_{eff}=0$
triggering, therefore, the condensation of the $\vec{k}=0^+$
fluctuations.

\section{Dynamical properties}

In order to investigate the dynamics, we have solved numerically
the coupled equations (\ref{1.12bis}), (\ref{1.18}) and
(\ref{1.19}) in a discretized three-dimensional Fourier space
using fourth-order Runge-Kutta method with adaptive step size
\cite{NumRec}. We have used a mesh of linear size $L=1000$, taking
$\lambda =1$ and the initial conditions $m(0)=-1$,
$C_{\perp}(k,0)=0$.

\begin{figure}
\begin{center}
  \includegraphics[width=75mm]{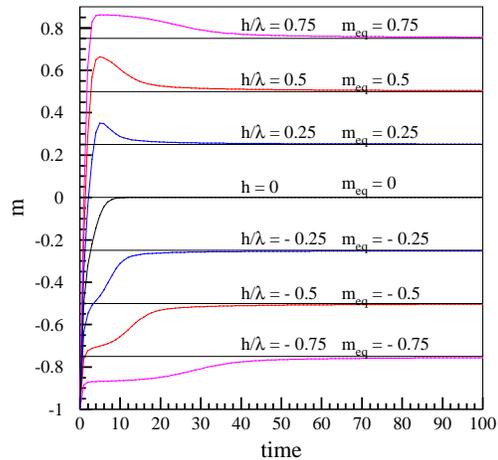}
  \caption{(Color online) Time evolution of the magnetization for different values of $h$,
   $\lambda=1$ and $T=0.25 \,T_C(h)$.}
  \label{magt}
\end{center}
\end{figure}

\begin{figure}
\begin{center}
  \includegraphics[width=75mm]{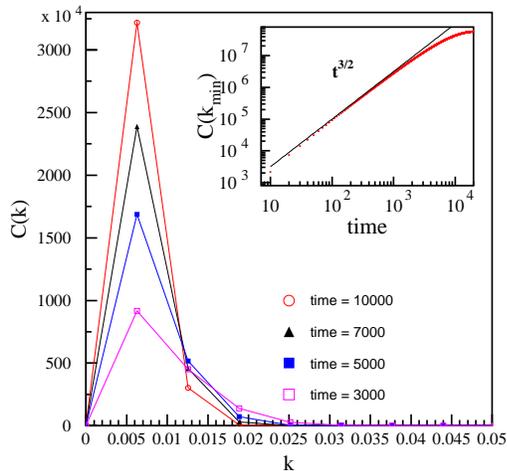}
  \caption{(Color online) Time evolution of the transverse structure factor $C_{\perp}(k,t)$
   for $h=0.5$, $\lambda=1$ and $T=0.25T_C(h)$. Inset:
    time evolution of the peak height $C_{\perp}(k_{min},t)$}.
  \label{ckt}
\end{center}
\end{figure}

Let us first consider $(T,h)$ in the region of phase separation, with $T<T_C(h)$.
Fig.~\ref{magt} and Fig.~\ref{ckt} illustrate quite well the considerations
made at the end of the previous section. The first one displays the evolution of
the magnetization for different values of $h$. After a fast transient
there is saturation to the equilibrium value
\be
m_{eq}= {h \over \lambda}
\label{3.1}
\ee
taking place from above or from below for $h$ positive or negative, respectively.
Apart from the details of the transient, this behavior of the magnetization agrees
with the prediction and results of Ref.~\cite{epl+FDGC+08} for the scalar model.

As explained above, when the magnetization gets stabilized at the value~(\ref{3.1}) by $h_{eff}=0$,
the growth of the Bragg peak is inevitable in order to equilibrate the system. This
is illustrated in Fig. \ref{ckt}.
In the late time regime the structure factor
is expected to obey a scaling form of the type~\cite{Bra95}
\begin{eqnarray}\label{scaling}
C_{\perp}(k, t) \sim L^{d}(t) F(k L(t)),
\end{eqnarray}
where $F(k L(t))$ is a scaling function and $L(t) \sim t^{1/z}$ with $z=2$, as appropriate
for phase ordering processes without conservation of the order parameter,
is a time dependent characteristic length~\cite{nota2}.
The same growth law $t^{1/2}$ was found in Ref~\cite{epl+FDGC+08} for the mean cluster size.
The inset of Fig.~\ref{ckt} shows that the height of the peak
follows quite well the power law
\be
C(k_{min}, t) \sim t^{3/2}.
\label{peak}
\ee
Actually, for times of order $10^3-10^4$ there appears a deviation
from the power law~(\ref{peak}). This is a finite size effect,
unavoidable in the numerical computation and not to be confused
with the equilibration of the magnetization, which is independent
of the size of the system (notice the huge difference in the time
scales). Therefore, considering the infinite system, we have the
interesting instance of two observables  in the same system, one
of which (the magnetization) equilibrates rather quickly, while
the other (the structure factor) does not reach equilibrium in any
time scale.

\begin{figure}
\begin{center}
  \includegraphics[width=75mm]{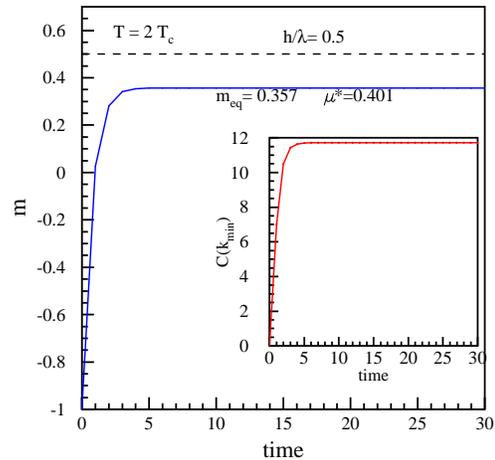}
  \caption{(Color online) Behavior of magnetization outside the coexistence region,
  with $h=0.5$, $\lambda=1$ and $T=2T_C(h)$.
    The dotted line represents the equilibrium value of magnetization for $T < T_C(h)$.
    Inset: time evolution of $C_{\perp}(k_{min},t)$.
}
  \label{mTlessTc}
\end{center}
\end{figure}

Conversely, if $T>T_c(h)$ there is a solution of the self-consistency relation without growth of the condensate
and both the magnetization and the structure factor reach equilibrium in the same
time scale. This is illustrated in
Fig. \ref{mTlessTc}, displaying the saturation of the magnetization
to the limit value $m_{eq}=\frac{h}{\lambda +\mu^*}$ with $\mu^* >0$, in agreement with
Eq.~(\ref{2.15}), while  the inset shows the saturation of the peak height to the equilibrium value
\be
T / (k_{min}^2 + \mu^*)
\label{00}
\ee
in agreement with Eq.~(\ref{Eq.16}).

The same qualitative behavior, that we have here illustrated in the $d=3$ case, is expected for any $d > 2$.
For $d=2$, due to the divergence in the denominator of Eq.~(\ref{2.10}), the critical temperature vanishes
squeezing the coexistence region of Fig. \ref{phase} onto the vertical axis at $T=0$. Therefore, in the $d=2$
case condensation of fluctuations can be obtained only for $T=0$ and $|h| \leq h_C$.

\section{Summary}
In this paper we have studied the static and dynamic properties of
a system described by a free energy functional with a short-range
ferromagnetic interaction and a long-range antiferromagnetic
interaction, in presence of an external uniform magnetic field.
The analysis has been carried out in the large $N$ limit. The
scalar counterpart of this model is the lattice-gas Hamiltonian
\cite{epl+FDGC+08} used to model the phenomenology of phase
separation occurring in the inner part of cell surface during
directional sensing. We have focussed on the phase ordering
process taking place below the critical temperature, even in the
in presence of the external magnetic field.

In particular, through the equations of motion
for the magnetization and the transverse structure factor, we have
highlighted how the competing interactions induce the self-tuning of the magnetization
within the phase coexistence region.
We have derived the phase diagram,
which depends on the magnetic field and the strength of the antiferromagnetic
coupling, showing that phase-separation is possible for a range of values of the external field.
Taking the large $N$ limit it has been possible
to derive analytically the dependence
of the critical temperature on the magnetic field and
on antiferromagnetic coupling $\lambda$. The phase diagram of Fig. \ref{phase} depicts in the $(T,h)$ plane
the phase coexistence regions for different values of $\lambda$.
The equilibrium value of magnetization $m_{eq}= h/\lambda$ is in agreement
with the prediction and with the results obtained in the Monte Carlo
simulation for a lattice gas system (Ref.~\cite{epl+FDGC+08}).
The dynamics shows that the antiferromagnetic coupling combines with the magnetization
to generate the effective magnetic $h_{eff}$ field, eventually vanishing within the coexistence
region. While the magnetization equilibrates very quickly, the structure factor
does not equilibrate on any time scale.

For $T<T_C(h)$ the relaxation process is characterized by the
growth of a condensate in transverse structure factor at the most
unstable wave vector $k=k_{min}$. The onset of condensation
signals the occurrence of a phase separation and corresponds to
domain coarsening in the scalar case. The late time behavior of
the structure factor is characterized by dynamical scaling and
power law growth of the peak, with the exponent $d/2$
characteristic of non conserved order parameter.

In conclusion, we have analyzed a model where the competition between
the short-range ferromagnetic and the long-range antiferromagnetic
interaction between two species leads to the phase separation
for a wide range of external field and temperature.
The occurrence of phase separation is a crucial intermediate step allowing for
the amplification of external field gradients, leading to directional sensing
as illustrated in Ref.~\cite{epl+FDGC+08}.

The general property of the free energy functional~(\ref{1.1}),
giving rise to phase coexistence through self-tuning, can be very useful
in other contexts characterized by the balance between
short range attraction, long-range repulsion and an overall external action.

{\bf Acknowledgements.} MZ acknowledges financial support from MURST through the
PRIN project 2007JHLPEZ {\it Fisica Statistica dei Sistemi Fortemente Correlati all'Equilibrio
e Fuori dall'Equilibrio: Risultati Esatti e Metodi di Teoria dei Campi}.


\begin{thebibliography}{4}
\bibitem{footnote} In practice such a final state
is never achieved, since an infinite system quenched below the critical point never equilibrates.
\bibitem{nota1} More precisely, $K_d$ is the $d$-dimensional solid angle over $(2\pi)^d$.
\bibitem{nota2} In the case of non linear systems with a genuine phase ordering process $L(t)$ is related to the size of coarsening domains.
\end{thebibliography}
\end{document}